# Error Analysis of CORDIC Processor with FPGA Implementation

Young-Man Kim

Electrical Engineering, Liberty University,
1971 University Blvd., Lynchburg, VA 24515, USA (ykim81@liberty.edu)

**Abstract:** The coordinate rotation digital computer (CORDIC) is a shift-add based fast computing algorithm which has been found in many digital signal processing (DSP) applications. In this paper, a detailed error analysis based on mean square error criteria and its implementation on FPGA is presented. Two considered error sources are an angle approximation error and a quantization error due to finite word length in fixed-point number system. The error bound and variance are discussed in theory. The CORDIC algorithm is implemented on FPGA using the Xilinx Zynq-7000 development board called ZedBoard®. Those results of theoretical error analysis are practically investigated by implementing it on actual FPGA board. In addition, Matlab is used to provide theoretical value as a baseline model by being set up in double-precision floating-point to compare it with the practical value of errors on FPGA implementation.

## 1. INTRODUCTION

The CORDIC processor is an elegant algorithm for fast processing using the simple shift-add operation and it is one of the most efficient computer arithmetic (Volder, 1959). Due to its simplicity and regularity, many researches have been pursued with this one and applied to various algorithms (Walther, 1971; Chen, 1972; Meher et al., 2009). Not only trigonometric function but also various transcendental functions could have been implemented. Recently, various applications were successfully verified in different engineering fields such as image processing, software radio in communications, DSP such as singular value decomposition, QR decomposition, fast Fourier transform, digital filter design, and even in machine learning (Sathyanarayana et al., 2007; Vankka, 2005; Ercegovac and Tang, 1990; Troya et al., 2008; Guil et al., 1995; Das and Banerjee, 2002; Parhi et al., 2000). Regarding accurate operation of the processor, it is necessary to employ error analysis based on the mean square error (MSE) value because the analysis is based on the variance and power concept thus, it can be applicable to many areas such as DSP, digital fault detection, system identification, machine learning etc. The propagated error variance through the pipelined CORDIC processor is researched and Matlab is used to simulate the result. Using Matlab, one can simulate the propagated error variance and its effect can be compared with that of actual implementation on FPGA. The error model which is implemented with Matlab is working as a baseline model. The numeric setting of Matlab is double-precision floating point because Matlab is running on PC, there is no limit in resources and by setting it in the most precise calculation mode, it can serve as an ideal baseline model (Matlab®, 2020). Recently popular field programmable gate array (FPGA) device requires an optimized approach to the design of the best area and power compensation (Meher et al., 2008).

However, the conventional analysis only provides the upper bound of the error (Hu 1992). Therefore, this paper addresses the issue and presents the MSE based error analysis result. In addition, the analysis is extended to a practical FPGA system of Xilinx Zynq-7000 ZedBoard®. The CORDIC algorithm is actually implemented on FPGA and the error is analyzed to investigate its characteristics in actual system. As for the FPGA development software, Xilinx Vivado 2020.1 version software is used because it fully supports from design to synthesis (ZedBoard, 2020; Vivado Design Suite, 2020; Y.-M. Kim, 2023). The paper is organized as follows: In section 2, some basis of CORDIC processor used for this research is introduced. In section 3, error covariance matrix and error propagation are derived and error bound is discussed. In section 4, the CORDIC processor is implemented on FPGA on Xilinx ZedBoard and the practical investigation of error characteristics is conducted and compared with a baseline model using Matlab. Conclusion is summarized in section 5.

## 2. BASIS OF CORDIC PROCESSOR

The basis of CORDIC processor is to compute a trigonometric function using vector rotation which is simply implemented in shift-add algorithm because the shift and addition can be completed in the fastest way in digital processor. There are basic three modes in CORDIC operation which was extended by (Walther, 1971) from the initial concept (Volder, 1959): circular, linear, and hyperbolic. In this research, we focus only on the circular mode. Each mode has two different operations: rotating and vectoring. The rotating mode is that the initial 2D vector ($[x(1) \quad y(1)]^T$) makes a rotation with a user-defined angle, $\theta$ and moves to new position ($[x_o(N) \quad y_o(N)]^T$). Compared with that, the vectoring mode is as follows: The initial 2D vector position is projected onto one of two dimensional axis to find the positional angle of the vector ($cos\theta, sin\theta$). The rotational angle is implemented as a series of incremental small angle which can be performed by shift-add operation:

$$\theta \cong \sum_{i=0}^{N-1} \sigma(i)\alpha(i) \qquad (1)$$

where $\sigma(i)$ is the direction of rotation (+1 for counter clockwise and -1 for clockwise), $\alpha(i)$ is an incremental angle ($\alpha(i) \triangleq 2^{-i}$) which can be easily implemented with the shift-to-the-right operation in digital processor, and $N$ is the number of rotation. We now connect this incremental angle for the rotation in 2D with positional vector, $(x(i), y(i), i = 0,1,\cdots, N-1)$. Finally, the complete equation for circular operation of CORDIC is:

$$\begin{bmatrix} x(i+1) \\ y(i+1) \end{bmatrix} = \frac{1}{\sqrt{1+2^{-2i}}} \begin{bmatrix} 1 & -\sigma(i)2^{-i} \\ \sigma(i)2^{-i} & 1 \end{bmatrix} \begin{bmatrix} x(i) \\ y(i) \end{bmatrix} \qquad (2)$$

$$\theta(i+1) = \theta(i) - \sigma(i)tan^{-1}(2^{-i}). \qquad (3)$$

The scaling factor at (2) can be defined this way:

$$k(i) \triangleq \sqrt{1+2^{-2i}}. \qquad (4)$$

The rotational matrix can be defined as follow:

$$P(i) \triangleq \begin{bmatrix} 1 & -\sigma(i)2^{-i} \\ \sigma(i)2^{-i} & 1 \end{bmatrix}. \qquad (5)$$

Using (4, 5), (2) can be rewritten as follows:

$$\begin{bmatrix} x(i+1) \\ y(i+1) \end{bmatrix} = k(i)^{-1} P(i) \begin{bmatrix} x(i) \\ y(i) \end{bmatrix}. \qquad (6)$$

If we expand the iterative multiplication of (6) from the first calculation to the $(N-1)^{th}$ iteration using (4, 5), we can describe it follows:

$$\begin{bmatrix} x(N) \\ y(N) \end{bmatrix} = (\prod_{i=0}^{N-1} k(i))^{-1} \prod_{i=0}^{N-1} P(i) \begin{bmatrix} x(0) \\ y(0) \end{bmatrix}. \qquad (7)$$

The total scaling can be denoted as below with a parameter, $K$:

$$K(i)^{-1} \triangleq (\prod_{i=0}^{N-1} k(i))^{-1}. \qquad (8)$$

The matrix propagation through $N$-times product can be denoted as follows:

$$B(i) \triangleq \prod_{j=i}^{N-1} P(j), B(N-1) = I_{2\times 2} \qquad (9)$$

where $I_{2\times 2}$ is 2 x 2 identity matrix.

We can rewrite (7) in concise way using (8, 9) as follows:

$$\begin{bmatrix} x(N) \\ y(N) \end{bmatrix} = K(i)^{-1} B(i) \begin{bmatrix} x(i) \\ y(i) \end{bmatrix}. \qquad (10)$$

The (10) shows the relationship between the final position and the initial position via the one-time scaling at last iteration step and the propagation through matrix product.

The scaling parameter, $K$, can be saved to a memory for fast processing because it only depends on the iteration. But, the parameter, $\sigma(i)$, changes as shown at every iteration. In vectoring mode, the final values, $(x(N), y(N))$, can be thought as $(cos\theta, sin\theta)$.

The following figure 1 shows the basis of further analysis of three sources of error in the subsequent sections.

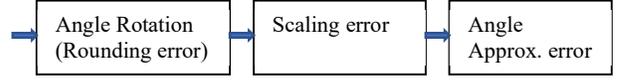

Fig. 1 Error sources

## 3. ERROR ANALYSIS

As you can see (3), the desired angle can never be achieved but, it exponentially approaches to it. Hence, one can easily understand the necessity of analysis of angle approximation error. The angle approximation error can be represented as follows:

$$\theta = \sum_{i=0}^{N-1} \sigma(i)\alpha(i) + \delta \qquad (11)$$

where $\delta$ represents the angle approximation error due to finite angle rotation which means the following.

$$\lim_{N\to\infty} \delta = 0. \qquad (12)$$

Using (12), we can describe it as follows:

$$\begin{bmatrix} x(\infty) \\ y(\infty) \end{bmatrix} = \begin{bmatrix} \cos\delta & -\sin\delta \\ \sin\delta & \cos\delta \end{bmatrix} \begin{bmatrix} x(N) \\ y(N) \end{bmatrix}. \qquad (13)$$

Thus, the angle approximation error, $e_a(N)$, which is defined as follows:

$$e_a(N) = \begin{bmatrix} x(\infty) \\ y(\infty) \end{bmatrix} - \begin{bmatrix} x(N) \\ y(N) \end{bmatrix}$$

$$= \begin{bmatrix} \cos\delta - 1 & -\sin\delta \\ \sin\delta & \cos\delta - 1 \end{bmatrix} \begin{bmatrix} x(N) \\ y(N) \end{bmatrix}. \qquad (14)$$

$e_a(N)$ is bounded as follows:

$$\|e_a(N)\| \leq \left\|\begin{bmatrix} \cos\delta - 1 & -\sin\delta \\ \sin\delta & \cos\delta - 1 \end{bmatrix}\right\|_2 \left\|\begin{bmatrix} x(N) \\ y(N) \end{bmatrix}\right\|$$

$$\Leftrightarrow \frac{\|e_a(N)\|}{\left\|\begin{bmatrix} x(N) \\ y(N) \end{bmatrix}\right\|} \leq \left\|\begin{bmatrix} \cos\delta - 1 & -\sin\delta \\ \sin\delta & \cos\delta - 1 \end{bmatrix}\right\|_2. \qquad (15)$$

At (15), we know $\left\|\begin{bmatrix} x(N) \\ y(N) \end{bmatrix}\right\| = 1$ because it goes through scaling process. Thus, the error bound is:

$$\|e_a(N)\| \leq \left\|\begin{bmatrix} \cos\delta - 1 & -\sin\delta \\ \sin\delta & \cos\delta - 1 \end{bmatrix}\right\|_2 \qquad (16)$$

$$= 2\sin\left|\frac{\delta}{2}\right| \cong |\delta|. \qquad (17)$$

The (17) is usually the case if the angle approximation error is very little. Thus, MSE of the scaling error is as follows:

*Theorem 1*: The MSE of angle approximation error can be described as follows:

$$E\|e_a(N)\|^2$$

$$= E\{[x(N) \quad y(N)]\begin{bmatrix} \cos\delta - 1 & \sin\delta \\ -\sin\delta & \cos\delta - 1 \end{bmatrix}\begin{bmatrix} \cos\delta - 1 & -\sin\delta \\ \sin\delta & \cos\delta - 1 \end{bmatrix}\begin{bmatrix} x(N) \\ y(N) \end{bmatrix}\}$$

$$= 4\sin^2\left(\frac{\delta}{2}\right) \qquad (18)$$

$$\cong \delta^2 \qquad (19)$$

*Proof*: The derivation from (17) to (18) is very obvious. (19) is usually the case if $\delta$ is very small.

Another source of error, $e_s$, is due to scaling. If the scaling data is saved at memory to be used, it goes through quantization process and introduces another error due to the finite word length. As previously mentioned, the scaling can be implemented in two ways. One is done at the same iteration when the rotational multiplication is calculated. The second one is that after the final $N^{th}$ iteration is done and simple one-time multiplication as shown in (10) is completed. For the fast processing, we assume here that the second one is the case. Based on that assumption, the following theorem tells us MSE due to the scaling error.

*Theorem 2*: If we assume the scaling is uniformly distributed over $[-2^{-b-1}, 2^{-b-1}]$ with b fractional bit system by simple multiplication after the last iteration, the mean and MSE of the scaling error is as follows:

$$E\{e_s\} = 0_{2\times 1}, \quad (20)$$

$$E\|e_s\|^2 = 2^{-2b}/12. \quad (21)$$

where $0_{2\times 1}$ denotes zero vector.

*Proof*: The (20), mean of the scaling error, is obvious with the given assumption. The (21) is MSE due to scaling error and the calculation is as follows:

$$E\|e_s\|^2 = \int_0^{2^{-b-1}} e_s^2 \left(\frac{2}{2\cdot 2^{-b-1}}\right) de_s = 2^{-2b}/12. \quad (22)$$

The third source of error is rounding error, $e_r$, which results from finite word length. In this research, we assume fixed-point with $b$ fractional number system which is usually denoted as $Q_{x.b}$. Because the direction of rotation can be either positive or negative, the binary number should be signed number. Here, we use 2's complement signed number system and the magnitude of the 2D vector is normalized to 1. Thus, the value of $x$ in $Q_{x.b}$ is 1 and the rounding error, $\epsilon$, is described as follows:

$$\epsilon = 2^{-b-1}. \quad (23)$$

Thus, the upper bound of 2D rounding error is as follows:

$$\|e_r\| \leq \sqrt{2}\epsilon. \quad (24)$$

The (24) shows the maximum rounding error which it can be. The worst case when the maximum total rounding error occurs is that at every iteration, maximum rounding error, $\sqrt{2}\epsilon$, happens. The total rounding error for total $N$ iterations is as follows:

$$e_r = \sum_{i=0}^{N-1} B(i) e_r(i) \quad (25)$$

where

$$e_r(i) = \begin{bmatrix} e_{rx}(i) \\ e_{ry}(i) \end{bmatrix} \quad (26)$$

and $|e_{rx}(i)| \leq \epsilon$, $|e_{ry}(i)| \leq \epsilon$.

*Theorem 3*: We assume that the rounding error is uniformly distributed over three cases of rounding error, $\{-\epsilon, 0, \epsilon\}$, then the mean and MSE of the rounding error is as follows:

$$E\{e_r(i)\} = 0_{2\times 1} \quad (27)$$

$$E\|e_r\|^2 = \frac{4\epsilon^2}{3}\{\sum_{j=0}^{N-1}[1 - \prod_{i=0}^{i=j}\sigma(i)2^{-i}]^2\}. \quad (28)$$

*Proof*: With the assumption of uniform distribution, the (27) is too obvious. Using (27), the MSE of $e_r$, $E\|e_r\|^2$, is as below.

$$E\|e_r\|^2 = E(e_r^T e_r)$$
$$= E\{[\sum_{i=0}^{N-1} B(i) e_r(i)]^T [\sum_{i=0}^{N-1} B(i) e_r(i)]\}. \quad (29)$$

After simple calculation with the replacement of (29) with (9) and (26), then, we arrive the following result:

$$E\|e_r\|^2 = \frac{4\epsilon^2}{3}\{\sum_{j=0}^{N-1}[1 - \prod_{i=0}^{i=j}\sigma(i)2^{-i}]^2\}.$$

Up to this point, we find three error sources and their MSE values. Because all three sources of error are independent, we can derive this conclusion about the total MSE of CORDIC operation described in this research as below.

*Theorem 4*: The total MSE of the CORDIC operation which performs one-time last scaling with rotation mode in $Q_{1.b}$ number system with $b$ fractional bits and rounding error, $\epsilon$ at (23) is as follows:

$$E\{\|e_{total}\|^2\} = E\|e_a(N)\|^2 + E\|e_s\|^2 + E\|e_r\|^2, \quad (30)$$

where each term is matched with (18), (21), and (28).

*Proof*: We assume that each source of all three errors is independent, thus (30) is obvious in the viewpoint of statistics (Gardner, 1990).

## 4. FPGA IMPLEMENTATION

The developed idea in previous section is implemented on Zynq-7000 FPGA which is at Xilinx ZedBoard to check its effectiveness. At first, Matlab is used to verify the idea. The Matlab numeric setting is done as double precision floating point number and it is used as a baseline model for this simulation which means it is infinitely correct value without precision limitation. The value is described as in the Angle [rad] column at the following table 1 below. Verilog is used to develop HDL code to implement the idea on FPGA. For the fractional bit, $b$, 15-bit is used. The radius of the rotation is 1 due to normalization from scaling. To represent clock and counter-clockwise rotational direction, one bit is used in two's complement format. Thus, we use $Q_{1.15}$ number system. The total bit number is 16. Sixteen pipeline is used for implementing the rotation. The $\theta$ value for rotation should be in the range of -0.9579 [rad] ~ +0.9579. For scaling which is conducted only one-time after the last iteration, the value of (8) and the value of $tan^{-1}(2^{-i})$ at (3) are saved in the memory for fast processing. Here, we calculate cosine and sine of the desired angle. The following table 1 shows the difference (error) representing the value of (17) is shown in the second and third column named 'cos_v-m_cos' and 'sin_v-m_sin' respectively. The values, 'cos_v' and 'sin_v', come from the FPGA implementation. The 16 pipeline structure is used thus, the depth is 16. The final hexa-value after rotation with 16 pipelines is shown at figure 2. Figure 3 shows actual implementation on Xilinx FPGA board. Figure 4 shows the utilization of FPGA slices from synthesis. Figure 5 and 6 show the power usage in detail which is 0.176 W. Figure 7 shows the timing analysis result and the max. achievable speed is:

$$1/4.337[ns] = 230.57\ [MHz]. \quad (31)$$

Table 1. Comparison of Matlab result with Actual FPGA implementation

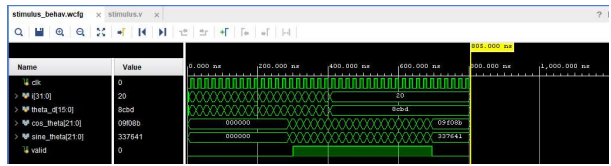

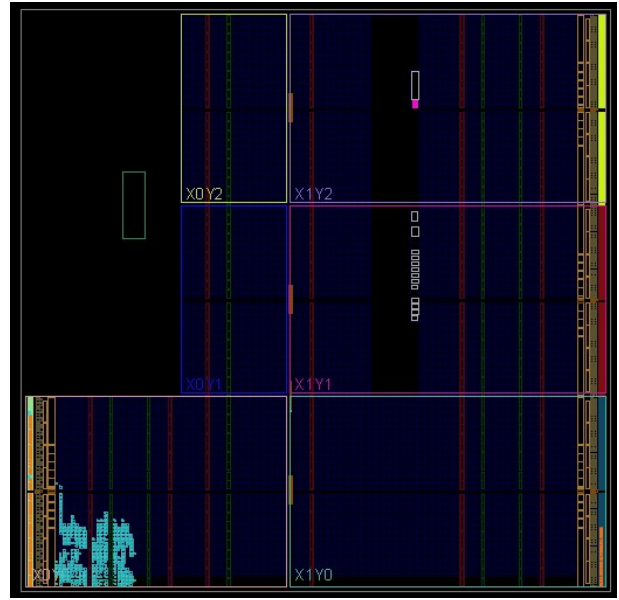

Fig. 3 Synthesis picture on FPGA

(790 LUTs, 851 F/Fs)

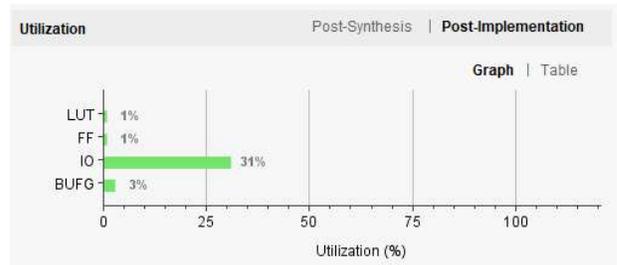

Fig. 4 Utilization report

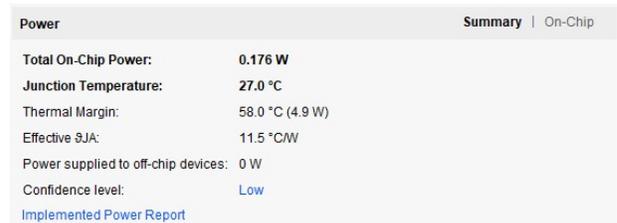

Fig. 2 Pipeline structure with rotational final hexa-value

Fig. 5 Power usage

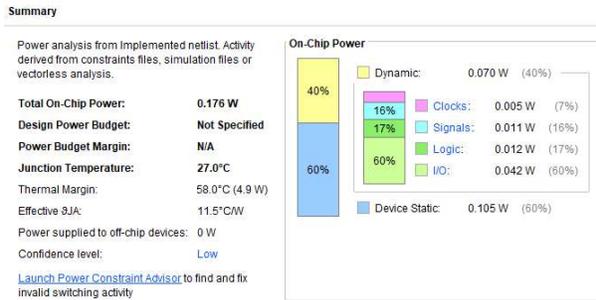

Fig. 6 Power usage in details

Fig. 7 Timing analysis result

## 5. CONCLUSION

In this research, we analyse error sources of CORDIC processor. The condition is that the final scaling for normalization is conducted only one-time at the last iteration for fast processing with fixed point system. The three error sources are rounding, scaling, and angle approximation. The error bound for each source is derived and then, MSE for each error is formulated. In addition, total MSE under the assumption of independence of each error source is explained. Using Matlab and Xilinx FPGA board, the effectiveness is verified. The detail synthesis and implementation on FPGA board is discussed.